\documentclass[a4paper,10pt,twocolumn]{article}

\usepackage{amsmath}
\usepackage{amssymb}
\usepackage[dvipdfmx]{graphicx}
\usepackage{ifthen}
\usepackage{cite}
\usepackage{verbatim}
\usepackage[usenames]{color}
\usepackage{slashbox}
\usepackage{url}
\usepackage{algorithm}
\usepackage{algorithmic}

\renewcommand{\algorithmicrequire}{\textbf{Input:}}
\renewcommand{\algorithmicensure}{\textbf{Output:}}

\begin{document}

\title{
	Eclipse Hashing: Alexandrov Compactification 
	and \\
	Hashing with Hyperspheres
	for Fast Similarity Search
}
\setcounter{footnote}{0}
\author{
Yui~Noma
\footnote{Software Systems Laboratories, FUJITSU LABORATORIES LTD.  1-1,
Kamikodanaka 4-chome, Nakahara-ku Kawasaki, 211-8588 Japan.}
\footnote{E-mail: \texttt{noma.yui@jp.fujitsu.com}}
\and
Makiko~Konoshima
\footnotemark[1]
\footnote{E-mail: \texttt{makiko@jp.fujitsu.com}
}
}
\date{}
\maketitle

\begin{abstract}


The similarity searches that use high-dimensional feature vectors
 consisting of a vast amount of data
 have a wide range of application.
One way of conducting 
 a fast similarity search
 is to 
 transform the feature vectors
 into binary vectors and perform the similarity search
 by using the Hamming distance.
Such a transformation is a hashing method,
 and the choice of hashing function is important.


Hashing methods using hyperplanes or 
 hyperspheres are proposed.
One study reported here 
 is inspired by Spherical LSH~\cite{SphericalHashing},
 and we use hypersperes to hash the feature vectors.


Our method, called Eclipse-hashing,
 performs a compactification of $\mathbb{R}^n$ by using the inverse stereographic projection,
 which is a kind of Alexandrov compactification.
By using Eclipse-hashing,
 one can obtain the hypersphere-hash function without explicitly using hyperspheres.
Hence, the number of nonlinear operations is reduced
 and the processing time of hashing becomes shorter.
Furthermore, we also show that as a result of improving the approximation accuracy,
 Eclipse-hashing is more accurate than hyperplane-hashing.


Keywords:
Locality-sensitive hashing,
Hypersphere,
Alexandrov compactification,
Inverse stereographic projection.

\end{abstract}


\section{Introduction}
\label{sec_introduction}


At present,
 there are great opportunities for those who can
 make good use of unstructured data,
 such as images, movies, and data captured by sensors.
Among the available utilities
 for dealing with unstructured data,
 similarity searches
 have a wide range of application.
For example, 
 similarity searches of images obtained by scanning biometric information
 are used for card-less micropayments
 and fraud detection.
Moreover, many similarity search methods
 work by extracting feature vectors
 from unstructured data.
These feature vectors reflect the complexity of the data 
 from which they were extracted
 and can have 
 hundreds or even thousands of dimension.


One also needs to retrieve data from tens millions or even billions of records.
In addition, to utilize the unstructured data,
 one has to associate it with existing structured data.
Furthermore, by taking the need for security into consideration,
 it is preferable to store such data in a database (DB).


If one does a similarity search naively,
 the processing time is proportional to the number of records.
Hence, a naive similarity search 
 conducted on a huge amount of records
 takes a very long time.
Because of this, 
 fast similarity search methods
 have been developed,
 and many index structures have been proposed for them.
Two such indexes are R-Tree~\cite{R-tree} and kd-tree~\cite{KDTree}.
Moreover, to utilize unstructured data,
 one needs an index structure that permits fast similarity searches 
 in a high-dimensional space.
However,
 such index structures
 have yet to be developed.


The system requirements of the actual applications of unstructured data
 can be met
 not only by the similarity searches but also by approximate-similarity searches.
Let us take
 the case of a card-less micropayment system that uses biometric information images
 as an example.
Here, a customer who wants to make a payment scans his biometric information image in a shop.
The image is sent to the micropayment system.
The system does a fast (approximate) similarity search of the images
 and finds two or more images that are similar to the query image
 from the database that stores one thousand images.
The similar images are sent to the biometric authentication engine, which is an accurate and heavy process,
 and the engine returns the person's ID.
The system then performs the payment process.
For such processes, 
 one needs a fast and accurate biometric authentication system
 (Please see the literature~\cite{SIMBA_KN} for details).
If one needs higher accuracy, 
 the biometric authentication engine and/or
 the approximate-similarity search
 have to be refined.


We took recent developments in memory technologies into consideration
 and paid attention to the following technique with high compatibility with in-memory DBs.
The method performs similarity searches by converting the feature vectors
 to bit-vectors and using the Hamming distance
 to represent the dissimilarities between the bit-vectors.
This method has been used, for example, in content-based image retrieval systems~\cite{
conf/cvpr/WangKC10, LSH_FS2, conf/icml/WangKC10, 512dimGISTdata}.
Since the Hamming distance calculation is
 much faster than
 the $L_2$ distance calculation of the feature vectors,
 a similarity search in Hamming distance space is very fast.
Furthermore,
 since the information in the feature vectors is represented 
 with bit-vectors that are smaller than the feature vectors,
 information on a large amount of unstructured data
 can be loaded into memory.
 However, since the hashing causes a loss of information,
 similarity searches using bit-vectors 
 are approximations of
 ones using feature vectors.
To improve the accuracy,
 one might try lengthening the bit-vectors.
However, a huge amount of very long bit-vectors cannot be stored 
 in memory.
Therefore, one has to devise a hashing method
 that produces an accurate approximate-similarity search
 with bit-vectors that are not too long.


The hashing methods that use hyperplanes are
 locality-sensitive hashings
 and are actively studied~\cite{
LSH_p-stable, LSH_IndykMotwani, LSH_SimHash,
LSH_RandomProjection, MCMC_Hashing, SIMBA_KN, MLH,
conf/cvpr/WangKC10, conf/icml/WangKC10, DensitySensitiveHashing, LSH_FS1}.
Interpreting these hashing methods from a geometrical viewpoint,
 we can see that they use a hyperplane to 
 divide the feature vector space
 into two regions,
 and assign $0$ or $1$
 to each region and the vectors contained in them.
The bit is determined by the orientation of the hyperplane.
By performing this operation many times, 
 they assign bit-vectors to the feature vectors.
In contract, in the case of a feature vector space is a $L_2$ metric space,
 it is natural to divide up the space by using hyperspheres
 instead of hyperplanes
 because the distance between feature vectors in a hypersphere
 is less than the diameter.
Therefore,
 it is expected that a hypersphere-hashing approximation
 would be more accurate than
 a hyperplane-hashing approximation.
A hashing method using hyperspheres was first proposed in~\cite{SphericalHashing}.


Hashing with a kernel has also been proposed~\cite{
 KLSH, 10.1109/CVPR.2012.6247912}.
We can use kernels to divide up a space with 
 complicated hypersurfaces,
 and this is expected to yield a more accurate approximation
 than that of hyperplane hashing.
Despite this,
 it noted in~\cite{SphericalHashing}
 that 
 many separate regions may have a same bit-vector.
Furthermore, it is hard to control 
 the generation of such regions.
Hence, the accuracy of the hashing with a kernel
 approximation would deteriorate in such cases.


The following two problems occur when we naively do the hashing with
 hyperspheres explained in detail in section~\ref{sec_motivation}.
The first problem is an effect caused by
 the existence of a shortcut through a neighborhood around the infinity.
The second problem is the occurrence of the destruction of localities.


We propose a new hashing scheme in this paper,
 called Eclipse-hashing.
Eclipse-hashing uses the inverse stereographic projection
 to solve the above-mentioned problems.

\section{Related work}
\label{sec_relatedWork}


Here, we explain methods that are related ours.
We will assume that the feature space is
 an $L_2$ distance space.

\subsection{Hashing with hyperplanes}
\label{sec_hashingWithHyperplane}


Let the feature space $V$ be an $N$-dimensional space, i.e., $\mathbb{R}^N$,
 and let $(x_1, x_2, \cdots, x_N)$ be the coordinates of a vector in $V$.
A hyperplane that crosses the origin of $V$ is called a linear hyperplane.
Otherwise, it is called an affine hyperplane.
Consider $B$ hyperplanes $\{ H^{(k)} \}_{k=1,\cdots, B}$ in $V$.
By letting $\{ \vec{n}^{(k)} \}_{k=1,\cdots, B}$ be the unit normal vectors of $H^{(k)}$
 and $\{ b^{(k)} \}_{k=1,\cdots, B}$ be the offset,
 the equation of $H^{(k)}$
 can be formulated as follows:
\begin{eqnarray}
	\phi^{(k)}(\vec{x})
	=
	\vec{n}^{(k)} \cdot \vec{x} + b^{(k)} .
\end{eqnarray} 
When the hyperplane $H^{(k)}$ is linear, $b^{(k)} = 0$.
The hash function defined by the $k$-th hyperplane
 is given\footnote{
 This hash function corresponds to the limit
 where $r\rightarrow \infty$ of the hash function in~\cite{LSH_p-stable}.
}:
\begin{eqnarray}
	h^{(k)}(\vec{x})
	=
	\begin{cases}
		1 & \text{if $\vec{n}^{(k)}\cdot\vec{x} +b^{(k)} >0 $,}\\
		0 & \text{otherwise.}
	\end{cases}
	\label{eq_hashingWithHPlane}
\end{eqnarray} 
Since this hash function is a linear operation, 
 it is fast.
The bit-vector assigned to $\vec{x}$
 is $\vec{h}(\vec{x}) = (h^{(1)}(\vec{x}), \cdots, h^{(B)}(\vec{x}))$.


The details of the hash functions depend on the normal vectors.
There are many ways to give the normal vectors:
 by sampling them randomly from a multi-dimensional
 standard normal distribution~\cite{LSH_RandomProjection},
 using PCA~\cite{PCAH},
 or by using the density of data~\cite{DensitySensitiveHashing}.
When labeled data are attached to the feature vectors,
 one can find the normal vector 
 by using supervised learning:
 S-LSH~\cite{SIMBA_KN}, M-LSH~\cite{MCMC_Hashing}, MLH~\cite{MLH}.

\begin{figure}[htbp]
 \begin{center}
  \includegraphics[scale=1.0]{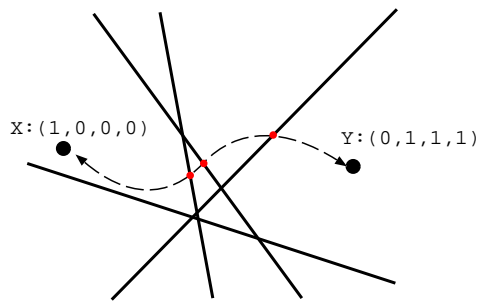}
 \end{center}
 \caption{Correspondence between the minimum number of intersections and the Hamming distance.
	The number of intersections is defined by
	the number of crossing points of the hyperplanes and the path connecting $X$ and $Y$.
	}
 \label{fig_intersectionNumHP}
\end{figure}

From a geometrical viewpoint,
 the operation of the hash functions $h^{(k)}$
 corresponds to
 using $B$ hyperplanes to divide up
 the feature space and assigning the bit-vectors to the feature vectors.
The Hamming distance between the bit-vectors converted from two feature vectors
 corresponds to the minimum of the number of intersections
 between the hyperplanes and the path connecting the two feature vectors.
We illustrate the correspondence in Fig.~\ref{fig_intersectionNumHP}.

\subsection{Hashing methods related to hyperspheres}
\label{sec_hashingWithHypersphere}


Let us explain the existing hashing methods that are related to hyperspheres.
In spherical hashing~\cite{SphericalHashing},
 one considers multiple hyperspheres in the feature space
 and assigns $1$ to the feature vectors if they are in the sphere
 and $0$, otherwise.
The similarities between the bit-vectors, however, are not computed with the Hamming distance,
 but with the spherical-Hamming distance defined by the following equation:
\begin{eqnarray}
	\mathrm{SphericalHammingDist}(b1,b2)  \nonumber \\
	:= |\mathrm{xor}(b1,b2)|/|\mathrm{and}(b1,b2)|,
	\label{eq_sphericalHamming}
\end{eqnarray} 
 where $|v|$ indicates counting the number of $1$s in the bit-vector $v$.
The spherical-Hamming distance, however,
 causes division by $0$ and the indeterminant $\frac{0}{0}$.
Therefore,
 the spherical-Hamming distance is not an appropriate way to determine the similarities
 between bit-vectors.


Another hashing method related to hyperspheres
 is Spherical LSH~\cite{SphericalLSH}.
This method is not a hashing method with hyperspheres
 but rather one with feature vectors
 located on a unit sphere in the feature space.
It uses high-dimensional regular polytopes.
The hash function assigns the ID of the vertex of the polytope
 to the feature vectors.
Hence, the hash value is not a bit-vector.

\section{Eclipse-hashing}

\subsection{Motivation}
\label{sec_motivation}


When the feature space is an $L_2$ distance space,
 hashing with hyperspheres~\cite{SphericalHashing}
 is a natural idea.
However, because of the discussion in 
 section~\ref{sec_hashingWithHypersphere},
 we use the Hamming distance
 to calculate the similarities between the bit-vectors
 instead of the spherical-Hamming distance.


Let us consider hashing with hyperspheres.
Suppose we have $B$ hyperspheres
 and denote the center of the $k$-th hypersphere as $\vec{p}_{\mathrm{HS}}^{(k)}$
 and the radius as $r_{\mathrm{HS}}^{(k)}$.
The naive hashing is as follows:
\begin{eqnarray}
	h_{\mathrm{HS}}^{(k)}(\vec{x})
	=
	\begin{cases}
		1 & \text{if $ \sum_{i=1}^B (x_i - (p_{\mathrm{HS}}^{(k)})_i)^2  < r_{\mathrm{HS}}^{(k)} $,}\\
		0 & \text{otherwise.}
	\end{cases}
	\label{eq_hashingWithHSphere}
\end{eqnarray}
Keeping in mind that the normal vectors are given randomly in~\cite{LSH_p-stable},
 we set the position and radius randomly in section~\ref{sec_experiment}.


Although it is not described in~\cite{SphericalHashing},
 we should point that there are two problems related to naive hashing with hyperspheres.

\begin{figure}[htbp]
 \begin{center}
  \includegraphics[scale=0.5]{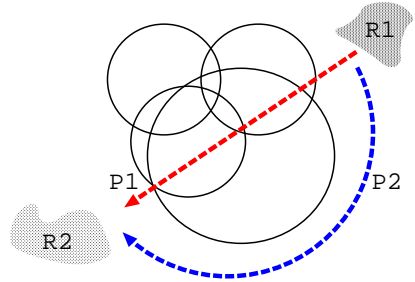}
 \end{center}
 \caption{Shortcut through the neighborhood of the infinity}
 \label{fig_shortcutThroughInfinityInPlane}
\end{figure}

-{\textit{Problem 1:
Effect of a shortcut through a neighborhood of the infinity.
}}
Let us consider the regions $R1$ and $R2$ in $V$,
 as shown in Fig.~\ref{fig_shortcutThroughInfinityInPlane}.
Since the bit-vectors assigned to them are the same,
 the Hamming distance between the points selected from each region is zero.
This means that
 path $P2$ is shorter than $P1$.
Hence, the accuracy of the approximation deteriorates.
$P2$ can be seen as a shortcut through the neighborhood of the infinity
 from the viewpoint of projective geometry.

\begin{figure}[htb]
 \begin{center}
  \includegraphics[scale=0.5]{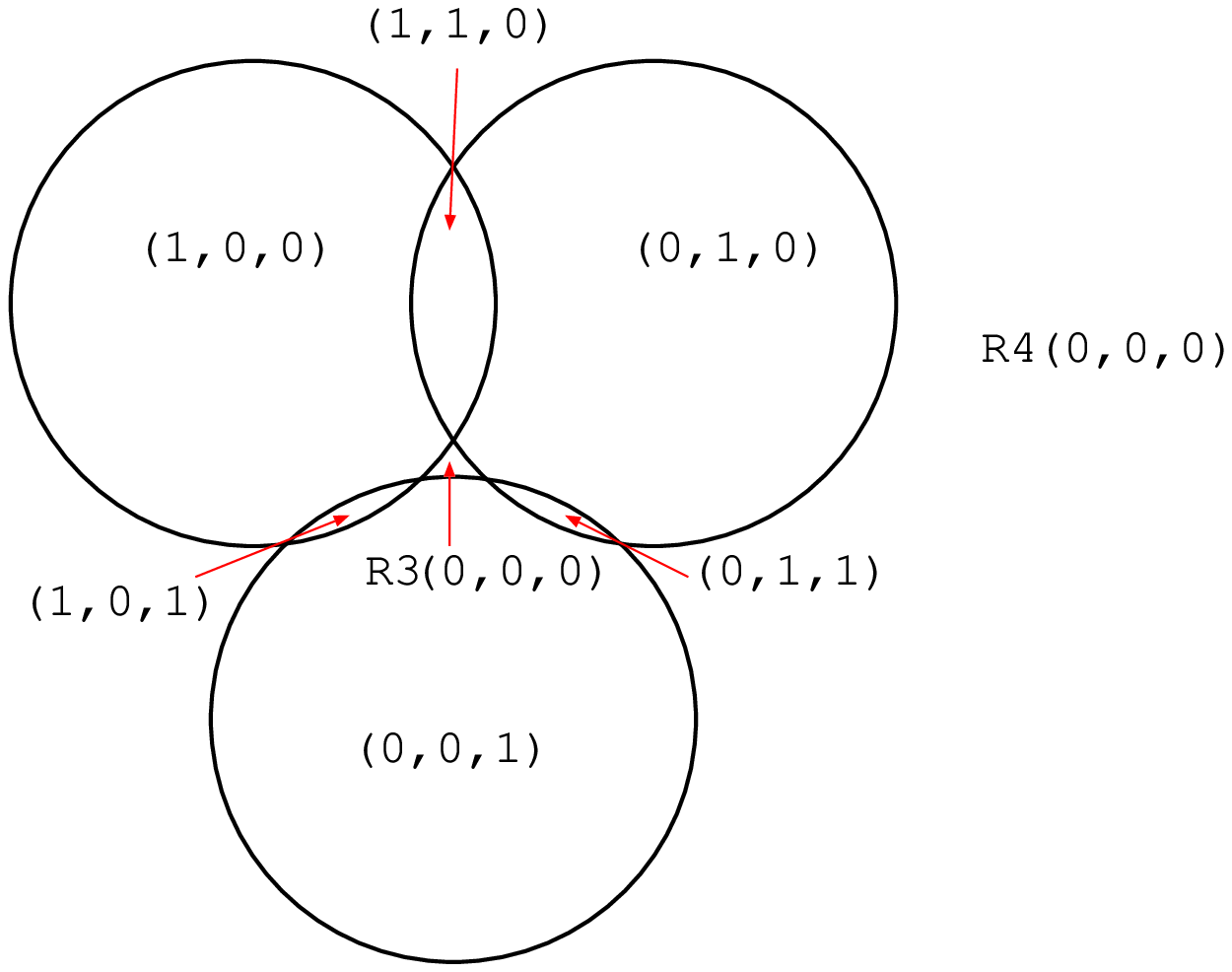}
 \end{center}
 \caption{Disconnectedness of a region having the bit-vector $(0,0,0)$}
 \label{fig_wormHoleInPlane}
 \begin{center}
  \includegraphics[scale=0.5]{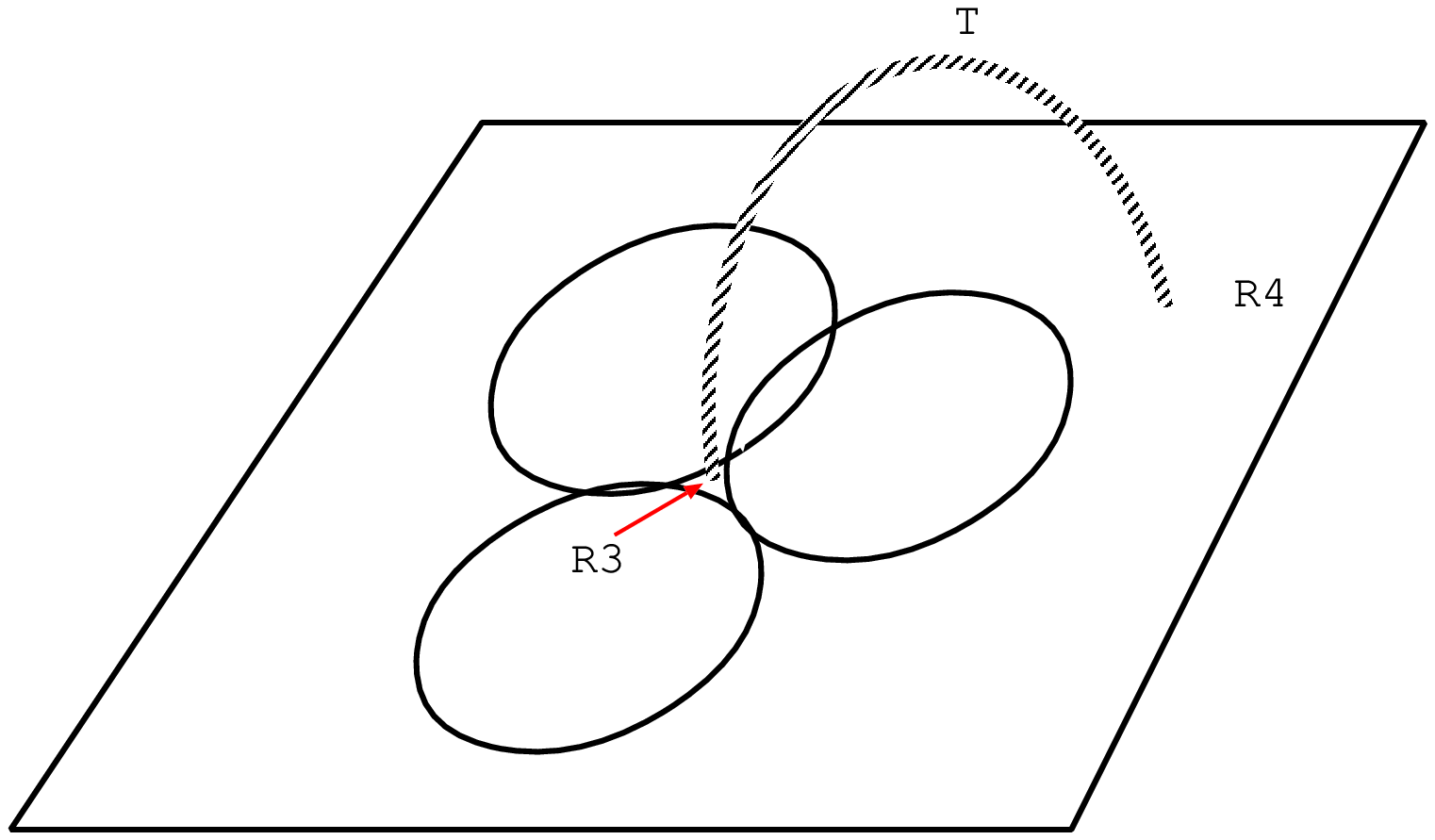}
 \end{center}
 \caption{
	Ambient space and tube connecting $R3$ and $R4$.
	}
 \label{fig_wormHoleInPlaneTube}
\end{figure}

-{\textit{Problem 2:
Disconnectivity of regions with the same bit-vector.
}}
Let us consider a case in which the feature space $V$ is two dimensional
 and the hashing is with three spheres,
 as shown in Fig.~\ref{fig_wormHoleInPlane}.
The bit-vectors assigned to the regions $R3$ and $R4$
 are the same, $(0,0,0)$.
In contrast,
 the region having the bit-vector $(0,0,0)$
 is disconnected.
In this case, the Hamming distance between 
 the bit-vectors corresponding to $a\in R3$ and $b\in R4$ 
 is zero.

However, any path connecting $a$ and $b$ 
 intersects the spheres.
Therefore, 
 the Hamming distance
 is different from
 the minimum number of intersections between the path and the spheres.
Because of this mismatch,
 one can not approximate the $L_2$ distance 
 by the Hamming distance.


We interpret this phenomenon as follows.
If a path connecting $a$ and $b$
 does not exist in the space $V$ and,
 except for its end points,
 is outside $V$, it does not intersect the spheres.
Let us consider a three-dimensional ambient space
 in which $V$ is located
 as shown in the Fig.~\ref{fig_wormHoleInPlaneTube}.
The ambient space can be seen as
 the direct product of the space $V$ and an extra dimension.
Furthermore,
 we consider the tube $T$ connecting $R3$ and $R4$,
 as shown in Fig.~\ref{fig_wormHoleInPlaneTube}.
When a path connecting $a$ and $b$ exists in this tube,
 the path does not intersect the spheres.
Therefore,
 by considering this ambient space and the tube
 the Hamming distance and
 the minimum number of intersections can be made to correspond.
We call $T$ a wormhole
 because it is similar to the wormholes
 predicted by general relativity.


In general, wormholes can exist when we partition a two-dimensional space
 with many spheres.
Similar things occur in the cases in which the dimension of the space is higher than two.
That is to say,
 when we divide up an $N$-dimensional space $V$
 with $N+1$ hyperspheres,
 a region corresponding to a bit-vector
 can be disconnected,
 and wormholes can exist.
Therefore,
 if we do the hashing with hyperspheres naively,
 the accuracy of the approximation deteriorates.
 

\subsection{Our contribution}


Our proposal solves the above-mentioned problems.
The key idea is the use of the inverse stereographic projection.
In the following
 we assume that
 the origin of $V$ and the mean vector of the feature vectors are the same.

\subsubsection{Inverse stereographic projection and hyperspheres}
\label{sec_inverseProjAndHypersphere}


Let us consider $\tilde{V} = \mathbb{R}^{N+1}$,
 and let $(\tilde{x}_1, \tilde{x}_2, \cdots, \tilde{x}_{N+1})$ be 
 the coordinate of $\tilde{V}$.
The following mapping is the so-called inverse stereographic projection
 $f^{-1}: V \rightarrow \tilde{V}$.
\begin{eqnarray}
  && f^{-1}(x_1, x_2, \cdots, x_N; d)  \nonumber \\
	&=& \left(
		\frac{2 d x_1}{d^2 + r^2}, 
		\cdots, \frac{2 d x_N}{d^2 + r^2},
		\frac{ -d^2 + r^2}{d^2 + r^2}
	\right),
	\label{eq_inverseStereoProjection}
\end{eqnarray} 
 where
 $r^2 := \sum_{i=1}^{N} x_i^2$, and $d$ is a given positive real number.
The image of $f^{-1}$ is a unit sphere $S$ whose center is the origin of $\tilde{V}$,
 except the north pole, $(0,0,\cdots, 1)$.
The function $f^{-1}$
 is the inverse 
 of the stereographic projection\footnote{
 The (inverse) stereographic projection is well known in mathematics.
} $f : S\setminus \{ (0,0,\cdots, 1)\} \rightarrow V$.

\begin{figure}[htbp]
 \begin{center}
  \includegraphics[scale=0.5]{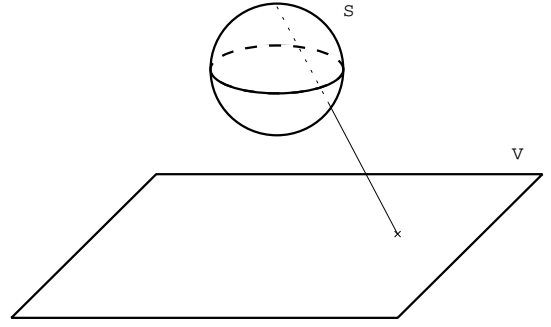}
 \end{center}
 \caption{Inverse stereographic projection}
 \label{fig_inverseStereoProjection}
\end{figure}

The dimensionality of $S$ is $N$.
Suppose we arrange
 $V$ at $\tilde{x}_{N+1}= -d+1$
 and consider rays
 whose initial point is the north pole of $S$;
 this mapping is a correspondence between
 the point where the ray and $S$ intersect
 and the intersection of the ray and $V$.
Figure~\ref{fig_inverseStereoProjection} shows the correspondence.
When we extend the domain of the mapping to contain the infinity of $V$,
 the image of the infinity is the north pole of $S$:
 i.e., $S$ can be identified as $V$ and the infinity, $S\cong V \cup \{ \infty \}$.
This identification is a kind of Alexandroff compactification.

\begin{figure*}[htb]
 \begin{center}
 \includegraphics[scale=0.5]{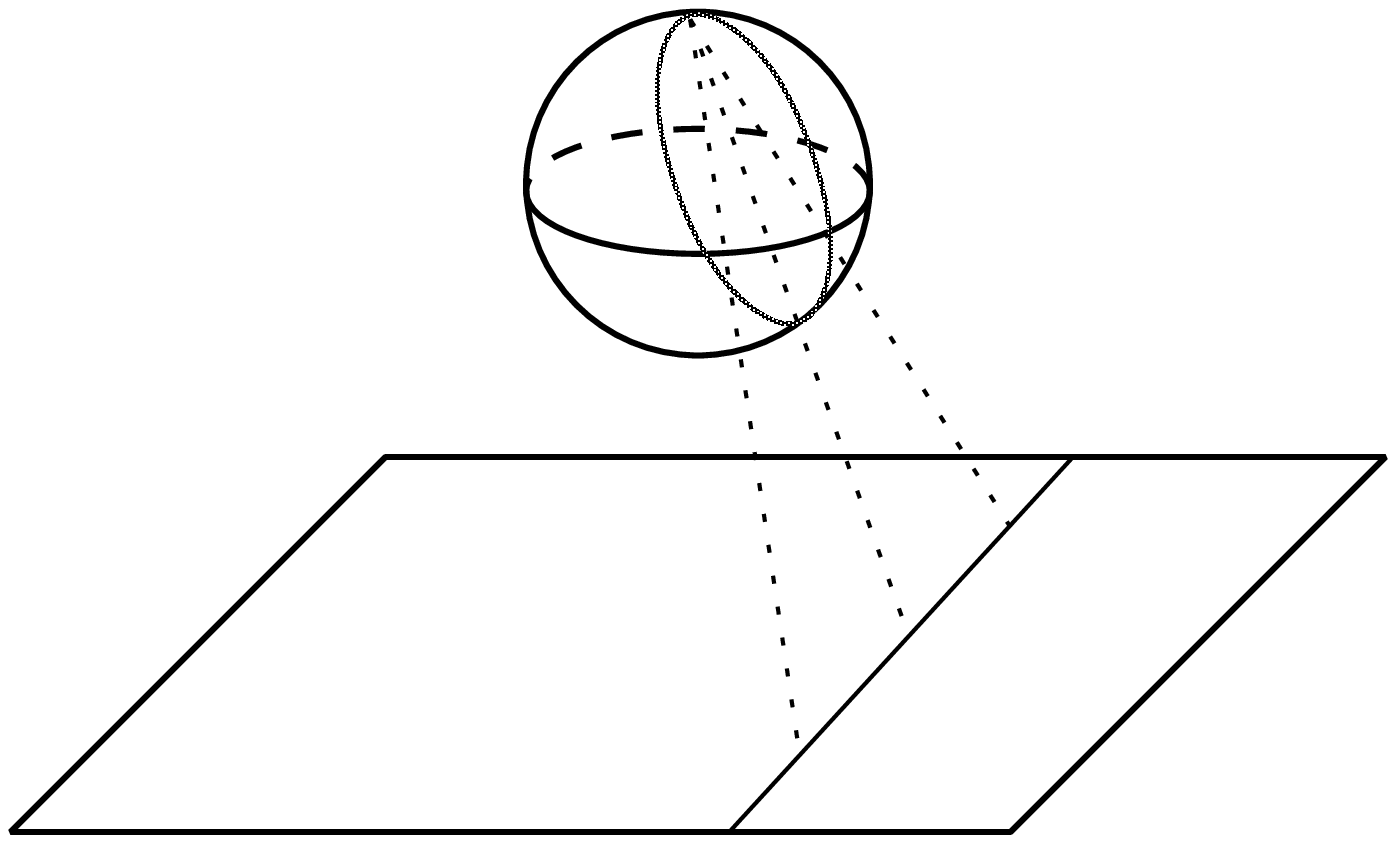}
  \includegraphics[scale=0.5]{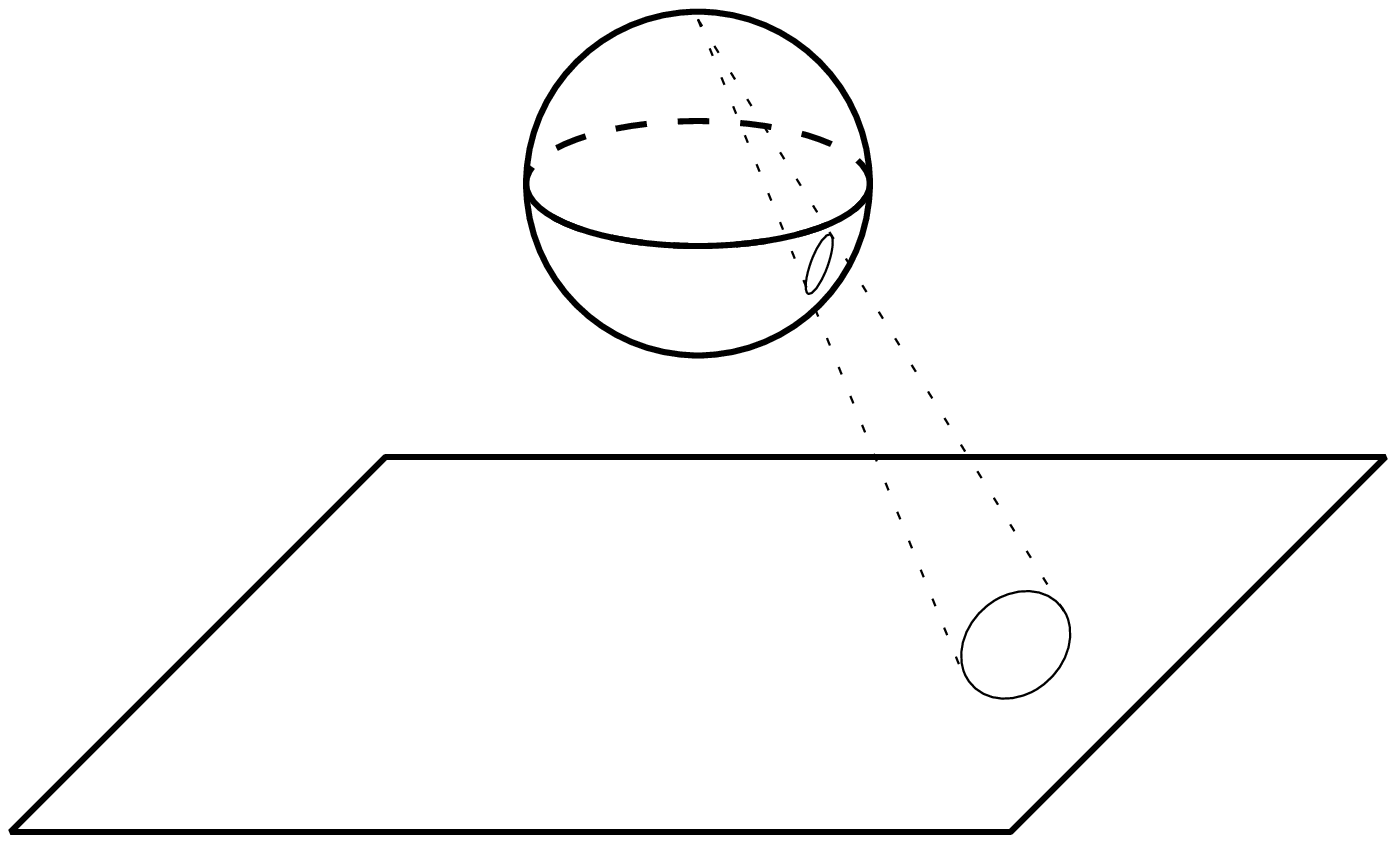}
 \end{center}
 \caption{
	The common set of a hyperplane in $\tilde{V}$ and $S$, and its image under $f$.
	The left figure shows the case where the common set contains the north pole,
		and the corresponding image is a hyperplane in $V$.
	The right figure shows the case where the common set does not contain the north pole,
		and the corresponding image is a hypersphere in $V$.
	}
 \label{fig_crosssectionSpherePlane}
\end{figure*}

Let us consider a partition
 of the unit sphere $S$
 by an affine hyperplane $\tilde{H}$ in $\tilde{V}$.
The dimensionality of $\tilde{H}$ is $N$ because it is a hyperplane
 in $(N+1)$-dimensional space $\tilde{V}$.
The equation of $\tilde{H}$ is:
\begin{eqnarray}
	\tilde{\phi}(\tilde{x}) = \vec{\tilde{n}} \cdot \vec{\tilde{x}} + \tilde{b} = 0 ,
	\label{eq_eqOfHyperPlaneInTildeV}
\end{eqnarray}
 where $\vec{\tilde{n}}= (\tilde{n}_1, \tilde{n}_2, \cdots, \tilde{n}_{N+1})$
 is the normal vector, 
 and $\tilde{b}$ is the offset.
If $\tilde{H}$ and $S$ have a common set,
 the image of the common set under the stereographic projection $f$
 is as follows:
\begin{eqnarray}
	0 &=& \tilde{\phi}(f^{-1}(x; d)) \nonumber \\
	&=&
	\sum_{i=1}^{N} \tilde{n}_i \frac{2d x_i}{d^2+r^2}
	+\tilde{n}_{N+1} \frac{-d^2+r^2}{d^2+r^2}
	+\tilde{b}
	\label{eq_intersection}
\end{eqnarray}
In the case of $\tilde{n}_{N+1} = - \tilde{b}$,
 we obtain the following equation
 from eq.(\ref{eq_intersection}).
\begin{eqnarray}
	\sum_{i=1}^N \tilde{n}_i x_i +d \tilde{b}=0 .
\end{eqnarray}
This is the equation of 
 an affine hyperplane in $V$.
In the case of $\tilde{n}_{N+1} \not= - \tilde{b}$,
 we obtain
\begin{eqnarray}
	&& \sum_{i=1}^N \left(
		x_i + \frac{d \tilde{n}_i}{\tilde{n}_{N+1}+\tilde{b}}
		\right)^2
		 \nonumber \\
		&= &
		\frac{d^2}{(\tilde{n}_{N+1}+\tilde{b})^2}
		\left(
			\sum_{i=1}^{N+1} \tilde{n}_i^2 - \tilde{b}^2
		\right)
		.
\end{eqnarray}
This is the equation of a hypersphere
 in $V$.
Especially
 in the case of $\vec{\tilde{n}} = (0,\cdots, 0,1)$ and $\tilde{b}=0$,
 the common set is the equator $E$ of $S$.
The image of $E$ under $f$ is a hypersphere,
 and the radius of the hypersphere is $d$.
From the above discussion,
 the image of the common set of $\tilde{H}$ and $S$
 under the stereographic projection
 is an affine hyperplane or a hypersphere.
Figure~\ref{fig_crosssectionSpherePlane}
 shows sketches of this situation.


Since the degree of freedom of a hyperplane in $\tilde{V}$
 is equal to that of
 a hypersphere or a hyperplane in $V$,
 there is a one-to-one correspondence
 between
 hyperspheres and hyperplanes in $V$
 and
 hyperplanes in $\tilde{V}$ whose common set with $S$ are not empty.
Furthermore,
 the two regions in $S$ separated by a hyperplane in $\tilde{V}$ 
 and the two regions in $V$ separated by the corresponding hypersphere or hyperplane
 coincide
 because $f$ and $f^{-1}$ are continuous.

\subsubsection{Solution to the problems}
\label{sec_solutionToProblems}

\begin{figure*}[tb]
 \begin{center}
  \includegraphics[scale=0.5]{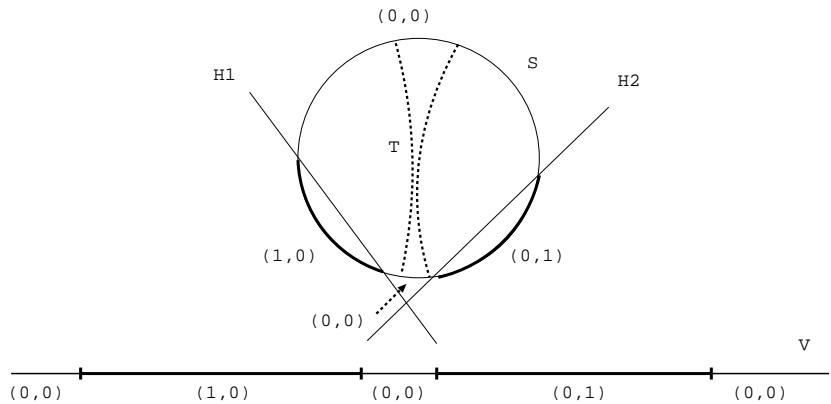}
 \end{center}
 \caption{
	Hashing with hyperspheres in a one-dimensional space and
	a wormhole existing inside $S$.
	}
 \label{fig_wormHoleIn12}
\end{figure*}


Since division of $S$ by hyperplanes
 and division of $V$ by hyperspheres
 correspond under $f^{-1}$,
 as we mentioned in section~\ref{sec_inverseProjAndHypersphere}, 
 hashing with hyperplanes in $\tilde{V}$
 after mapping the feature vectors to $S$ by $f^{-1}$
 corresponds
 to hashing with the corresponding hyperspheres in $V$.\footnote{
We need to invert the bits obtained from eq.(\ref{eq_hashingWithHPlane})
 in order to equate them to
 the ones  
 obtained from the composition of $f^{-1}$ and the hash function with hyperplanes in $\tilde{V}$.
}
By using this correspondence,
 we propose the following solutions
 to the problems mentioned in section~\ref{sec_motivation}.


-{\textit{
Solution 1: 
}}
There are two factors contributing to the deterioration
 in accuracy of the approximation
 affected by the shortcut through the neighborhood of the infinity.
The first factor is that
 the feature vectors that are far away from the origin of $V$
 are mapped to the neighborhood of the north pole of $S$,
 and the mapped points are near to each other in $\tilde{V}$.
The second factor is that
 unless the hyperplanes in $\tilde{V}$ do not 
 cross the neighborhood of the north pole of $S$,
 the Hamming distance of the bit-vectors assigned to the points near the north pole
 will be small.


If either of the two factors is resolved,
 the accuracy of the approximation may be improved.
On the one hand,
 even if the features are mapped near the north pole,
 the Hamming distances of the corresponding bit-vectors will be large 
 if many hyperplanes cross the neighborhood of the north pole.
On the other hand,
 even if the hyperplanes do not cross the neighborhood of the north pole,
 the Hamming distances of the corresponding bit-vectors
 are small if the feature vectors are not mapped to the north pole.
In the former case,
 since the hyperplanes cross near the pole,
 their image in $V$ under $f$
 are hyperspheres with a large radius that can be approximated by hyperplanes in $V$.
Hence, we chose the latter resolution.
Since the parameter $d$ of $f^{-1}$,
 corresponds the the radius of $f(E)$ as mentioned in section~\ref{sec_inverseProjAndHypersphere}, 
 we choose a large $d$
 in order to map the feature vectors 
 to regions away from the pole.


-{\textit{
Solution 2: 
}}
As we discussed in~\ref{sec_motivation},
 the accuracy of the approximation
 deteriorates because of wormholes.
Hence, if we could suppress
 the generation of wormholes
 or
 guarantee their non-existence,
 the accuracy would improve.
To be able to do this, though,
 we need to understand
 the generation process.


Let us consider a one-dimensional feature space to simplify
 the discussion.
In the case of $N=1$,
 the region surrounded by a hypersphere is a closed segment.
Figure~\ref{fig_wormHoleIn12}
 shows an example of two hyperspheres dividing up $V$.
The region having the bit-vector $(0,0)$
 are disconnected.
Therefore, a wormhole certainly exists.
The region having the bit-vector $(0,0)$ is mapped by $f^{-1}$
 to the two regions,
 the neighborhoods of the south and north poles.
As illustrated in Fig.~\ref{fig_wormHoleIn12},
 we can connect the two regions by putting a tube $T$
 through the inside of $S$.
The tube $T$ is the wormhole.
By looking at figure~\ref{fig_wormHoleIn12},
 we can understand the following:
 the wormhole is generated 
 because 
 the intersection of $H1$ and $H2$
 exists outside of $S$.
Hence, if the intersection 
 exists inside or at $S$,
 a wormhole would not be generated.



Although the above discussion is for the case of $N=1$,
 wormholes may exist if the dimension of $V$ is greater than one:
 they may be generated by $N+1$ hyperspheres in a $N$-dimensional space.
However, if the intersection point of the $N+1$ hyperplanes in $\tilde{V}$
 exists inside or at $S$,
 a wormhole would not
 be formed as a result of using $N+1$ hyperplanes.
When the length of the bit-vector is $B$,
 there is no wormhole
 if all the intersections of $B$ hyperplanes in $\tilde{V}$
 are inside or at $S$.
The number of intersections is $\frac{B!}{N!(B-N)!}$,
 since
 there is one intersection for each combination of $N+1$ hyperplanes.
Hence, when $N$ and $B$ are large,
 it is very heavy task
 to check all the intersection are inside or at $S$
 in order to guarantee that wormholes do not exist.


One of the simplest solutions 
 to reducing the size of the task
 is to require the following condition: 
 all hyperplanes in $\tilde{V}$ cross a point that exists inside or at $S$.
We call such a point a common intersection and denote it by $C$.
Please note that
 all hyperplanes crossing $C$ have a common set with $S$.

\subsubsection{Eclipse-hashing}


Based on the discussion in section~\ref{sec_inverseProjAndHypersphere} and ~\ref{sec_solutionToProblems},
 we can describe the Eclipse-hashing as follows.
Map the feature vectors in $V$ to $S$ by using 
 the inverse stereographic projection (eq.(\ref{eq_inverseStereoProjection})).
Select a common intersection $C$ inside or at $S$.
Chose the normal vectors of $B$ hyperplanes in $\tilde{V}$
 that cross $C$.
Hash the mapped feature vectors by using the hashing method with
 hyperplanes.\footnote{
 By regarding the space $V$ as a flat earth,
  the sphere $S$ can be seen as the moon in the universe.
 Dividing up the moon and assigning $0$ or $1$ to each region
  reminded us of a lunar eclipse.
 Hence, we decided to call this method Eclipse-hashing. 
}
The hash function is:
\begin{eqnarray}
	\tilde{h}^{(k)}(\vec{x})
	=
	\begin{cases}
		1 & \text{if $\vec{\tilde{n}}^{(k)}\cdot (f^{-1}(\vec{x}; d) -\vec{C}) >0 $,}\\
		0 & \text{otherwise ,}
	\end{cases}
	\label{eq_eclipseHashing}
\end{eqnarray}
 where $\vec{C}$ is the position vector of $C$.


Let $W$ be a $B\times (N+1)$ matrix whose $k$-th row vector is
 the normal vector $\vec{\tilde{n}}^{(k)}$
 of the $k$-th hyperplane.
The pseudo-code of Eclipse-hashing 
 is listed in Algorithm~\ref{algo_eclipseHashing}.
\begin{algorithm}
\caption{Eclipse-hashing}
\label{algo_eclipseHashing}
\begin{algorithmic}
	\REQUIRE $\vec{x}, W, \vec{C}, d$
	\STATE $\vec{\tilde{x}} \leftarrow f^{-1}(\vec{x}; d)$
	\STATE $\vec{\tilde{x}} \leftarrow \vec{\tilde{x}} -\vec{C}$
	\STATE $b \leftarrow \frac{1}{2} ( 1+sgn( W  \vec{\tilde{x}}))$
	\ENSURE $b$
\end{algorithmic}
\end{algorithm}


The parameters in Eclipse hashing
 are $d$ in $f^{-1}$, the common intersection $C$,
 and the normal vectors of the hyperplanes in $\tilde{V}$.
One can determine the normal vectors according to the purpose.
Some of the determination methods are
 unsupervised learning like in the literature~\cite{LSH_RandomProjection}
 or supervised learning ~\cite{SIMBA_KN, MCMC_Hashing, MLH}.
The parameters $d$ and $C$
 should be appropriately adjusted in order to 
 solve the problems, as mentioned in section~\ref{sec_solutionToProblems}.
We show in section~\ref{sec_experiment}
 that 
 the accuracy of the approximation can be improved by adjusting these parameters.

%
%
%

All the hash functions treated in Eclipse-hashing
 are expressed with hyperspheres.
However, Eclipse-hashing has the following advantages:
\begin{itemize}
\item
By simply setting a common intersection,
 we can guarantee that wormholes will not exist.

\item
By mapping the feature vectors to $S$, and translating
 them as $C$ becomes the origin,
 all the hyperplanes that we are considering
 cross the origin of $\tilde{V}$.
Therefore,
 we can apply existing supervised learning methods,
 for example ~\cite{MCMC_Hashing}~\cite{SIMBA_KN}~\cite{MLH},
 to determine the normal vectors of hyperplanes in $\tilde{V}$.

\end{itemize}

\section{Experiments}
\label{sec_experiment}


We measured the processing time and the accuracy of the approximation
 of Eclipse-hashing.
Furthermore,
 we measured the performances of the following hashing methods:
 $\mathrm{LH}$; hashing with linear hyperplanes in $V$,
 $\mathrm{AH}$; hashing with affine hyperplanes in $V$,
 and $\mathrm{HS}$; naive hashing with hyperspheres in $V$.
In the following, we abbreviate Eclipse-hashing as $\mathrm{EH}$.


In order to simplify the discussion,
 we set $\vec{C} = (0,0,\cdots, c)$.
The condition where $c\in [-1,1]$ must be imposed to ensure that
 the common intersection is inside or at $S$.

\subsection{Processing time of hashing}
\label{sec_procTime}


Here,
 we report the processing times of hashing of all the methods.
The experiment's environment was as follows.
The CPU was an Intel Xeon X5680 3.3GHz, and
 the size of the main memory was 32.0GB.
Each method was implemented using C++,
 each process was a single thread,
 and the linear algebra library Eigen~\cite{eigenweb} was used.
We used the Time Stamp Counter to measure the processing times.


The feature vectors were artificially made data
 that were sampled 
 from the 512-dimensional standard normal distribution.
The number of the feature vectors was 10,000.
Since the processing time of $\mathrm{EH}$ does not depend on the parameters $c$ and $d$,
 we set $c=0$ and $d=1$.
In addition,
 the normal vectors and the offsets of the hyperplanes
 and the centers and the radii of the hyperspheres
 were sampled 
 from the 512-dimensional standard normal distribution,
 since they do not affect the processing times
 of all the methods.

\begin{figure}[htbp]
 \begin{center}
  \includegraphics[scale=0.5]{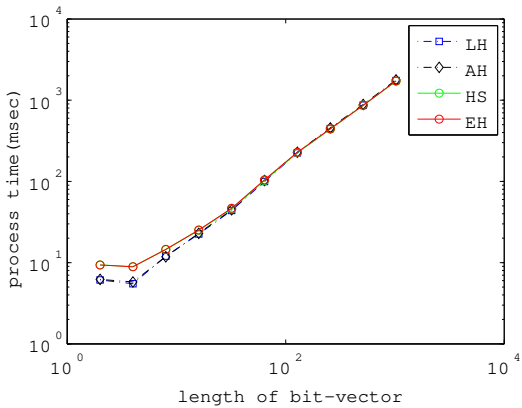}
 \end{center}
 \caption{
	Hashing processing time
	for 512-dimensional artificial data.
	}
 \label{fig_procTimeLogLogdim512_2}
\end{figure}

Figure~\ref{fig_procTimeLogLogdim512_2}
 shows the processing time of hashing with each method
 versus the length of the bit-vectors.
The figure reveals that
 the processing time of $\mathrm{EH}$ is similar to that of $\mathrm{HS}$
 and approaches those of $\mathrm{LH}$ and $\mathrm{AH}$
 as the bit-vectors becomes longer.
For example,
 when the length of the bit-vectors is 1,024,
 hashing with $\mathrm{EH}$ is about four times as fast as 
 hashing with $\mathrm{HS}$.

\subsection{Accuracy of approximation}
\label{sec_accuracy}

\subsubsection{Experimental data sets}


The data sets used in the experiments were
 MNIST~\cite{MNIST}, which is a set of the raw image data of hand-written digits,
 LabelMe, which is a set of Gist feature vectors~\cite{GIST} extracted from image data
 that was published in reference~\cite{512dimGISTdata},
 and a set of artificial data that were sampled 
 from the 512-dimensional standard normal distribution.
Each data set was partitioned into two sets,
 one, the set of the data in the database,
 the other,
 the set of the data used for the queries.
We shifted the feature vectors so that the mean vectors of the data sets
 were zero vectors.
The data sets are summarized in
 Table~\ref{table_quantOfDataset}.
The length of the bit-vectors was 1,024.
\begin{table*}[htb]
	\begin{center}
	\caption{Quantities of the data sets}
	\begin{tabular}{l|r|r|r}
		\hline
		\hline
		\backslashbox{
		Parameter
		}{Data set} & MNIST & LabelMe & Artificial data  \\
		\hline
		Number of data for records
			& 60,000 & 11,000 & 10,000 \\
		Number of data for queries
		  & 10,000  & 11,000 &  1,000 \\
		Dimensionality
		  & 784    & 512 &  512 \\
		\hline
	\end{tabular}
	\label{table_quantOfDataset}
	\end{center}
\end{table*}

\subsubsection{Details of the experiment}


%
We used the mean of recalls as an indicator 
 to measure the accuracy of the approximation.
Let $Z(q, k)$ be the $k$-nearest neighbors for a query $q$.
By converting the feature vectors with each method,
 we denote $W(q, k; \mathrm{Meth})$ as the $k$-nearest neighbors for a query $q$
 in the Hamming space,
 where $\mathrm{Meth} = \mathrm{LH}, \mathrm{AH}, \mathrm{HS}, \mathrm{EH}$.
The indicator is
\begin{eqnarray}
	&& \mathrm{Recall}(k; \mathrm{Meth}) := \nonumber \\
		&&
	\frac{1}{\#(Q)}
	\sum_{q\in Q}
	\frac{\#(Z(q, k) \cap W(q, k; \mathrm{Meth}))}
	{k},
\end{eqnarray}
 where $Q$ is the set of queries.
In the following,
 we equated $k$ to the $1\%$ of the number of the record data.


The hyperplanes and hyperspheres for each method were set as follows.
The normal vectors of the hyperplanes for $\mathrm{EH}$
 were sampled from the $(N+1)$-dimensional standard normal distribution.
The normal vectors of the hyperplanes for $\mathrm{LH}$
 were sampled from the $N$-dimensional standard normal distribution.
For $\mathrm{AH}$,
 the normal vectors of the hyperplanes
 were sampled from the $N$-dimensional standard normal distribution,
 and the offsets of the hyperplanes 
 were sampled from a uniform distribution whose support was $[0,1]$.
The centers of the hyperspheres for $\mathrm{HS}$
 were sampled from the $N$-dimensional standard normal distribution,
 and the radii of the hyperspheres 
 were $\sqrt{N}$ times the absolute values of values sampled
 from the standard normal distribution.\footnote{
 The mean distance between the origin and values sampled from
  the $N$-dimensional standard normal distribution
  is about $\sqrt{N}$.
 Hence, the chosen hyperspheres partition the space near the origin of $V$.
}

\subsubsection{Results of the experiment}
\label{sec_accuracyExpResult}

\begin{figure}[htbp]
 \begin{center}
  \includegraphics[scale=0.5]{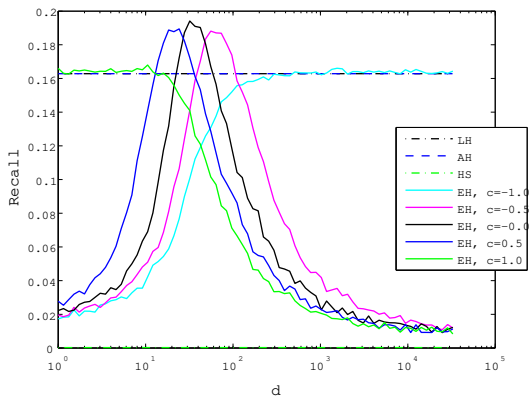}
 \end{center}
 \caption{
	Recall of each method
	on the artificial data.
	}
 \label{fig_RecallEcHAHSdim512sigma1.0k100bit1024Formated}
\end{figure}

Fig.~\ref{fig_RecallEcHAHSdim512sigma1.0k100bit1024Formated} shows
 $\mathrm{Recall}$
 of each method for the artificial data.
Since $\mathrm{Recall}(k; \mathrm{LH}) $,
 $\mathrm{Recall}(k; \mathrm{AH}) $,
 and $\mathrm{Recall}(k; \mathrm{HS})$ do not depend on $d$,
 they are drawn as straight lines.
$\mathrm{Recall}(k; \mathrm{HS})$ was close to zero.
Hence, $\mathrm{HS}$ had very bad accuracy on this data set.
$\mathrm{Recall}(k; \mathrm{AH}) $ and
 $\mathrm{Recall}(k; \mathrm{LH}) $ were almost the same.
Furthermore,
 $\mathrm{Recall}(k; \mathrm{EH})$ was
 greater than
 $\mathrm{Recall}(k; \mathrm{LH})$ and $\mathrm{Recall}(k; \mathrm{AH})$
 when $d$ was about $30$ and $c$ was about $0.0$.
This means that the accuracy of $\mathrm{EH}$
 is higher than $\mathrm{LH}$ and $\mathrm{AH}$
 with an appropriate choice of parameters.

\begin{figure*}[htbp]
 \begin{center}
  \includegraphics[scale=0.5]{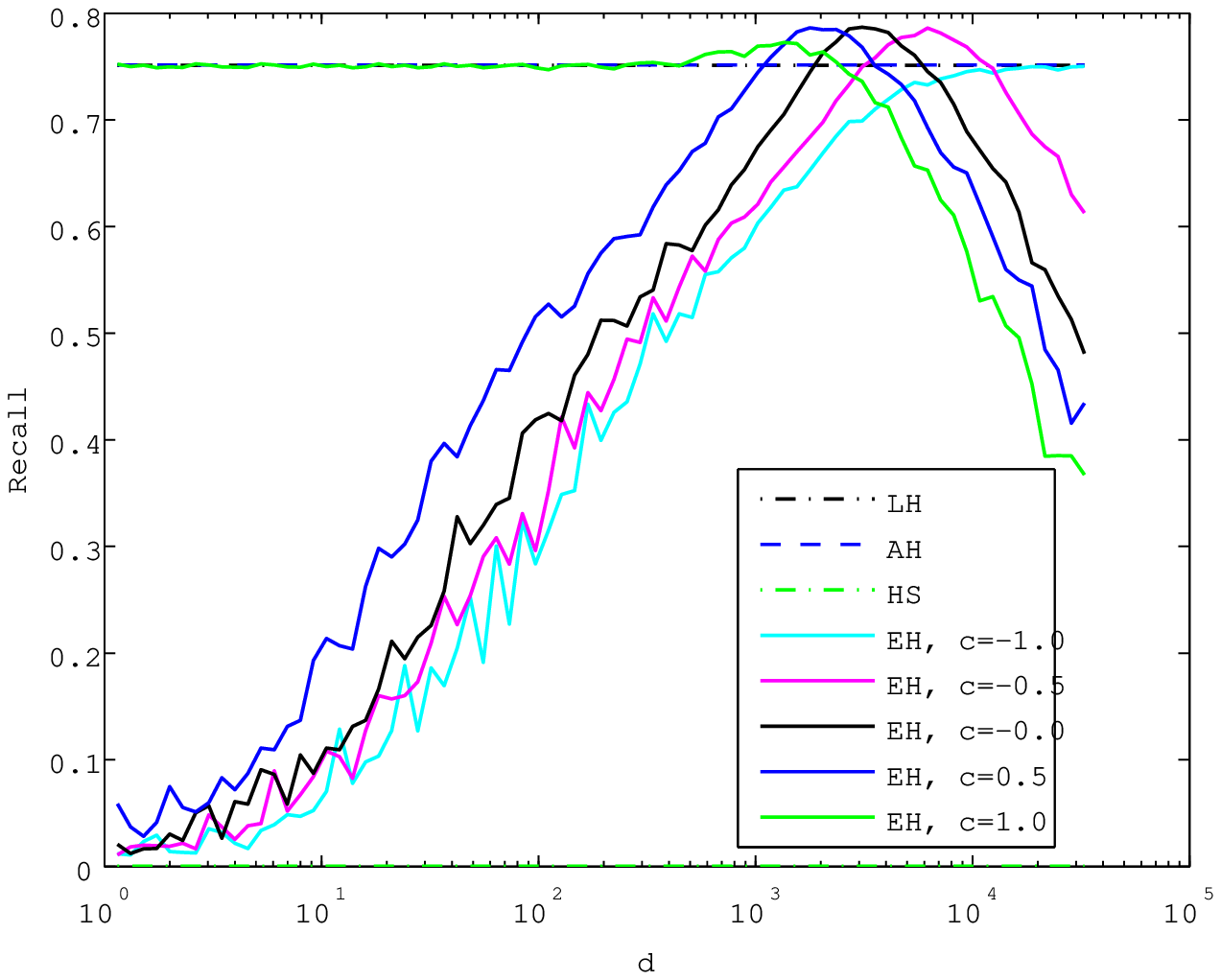}
  \includegraphics[scale=0.5]{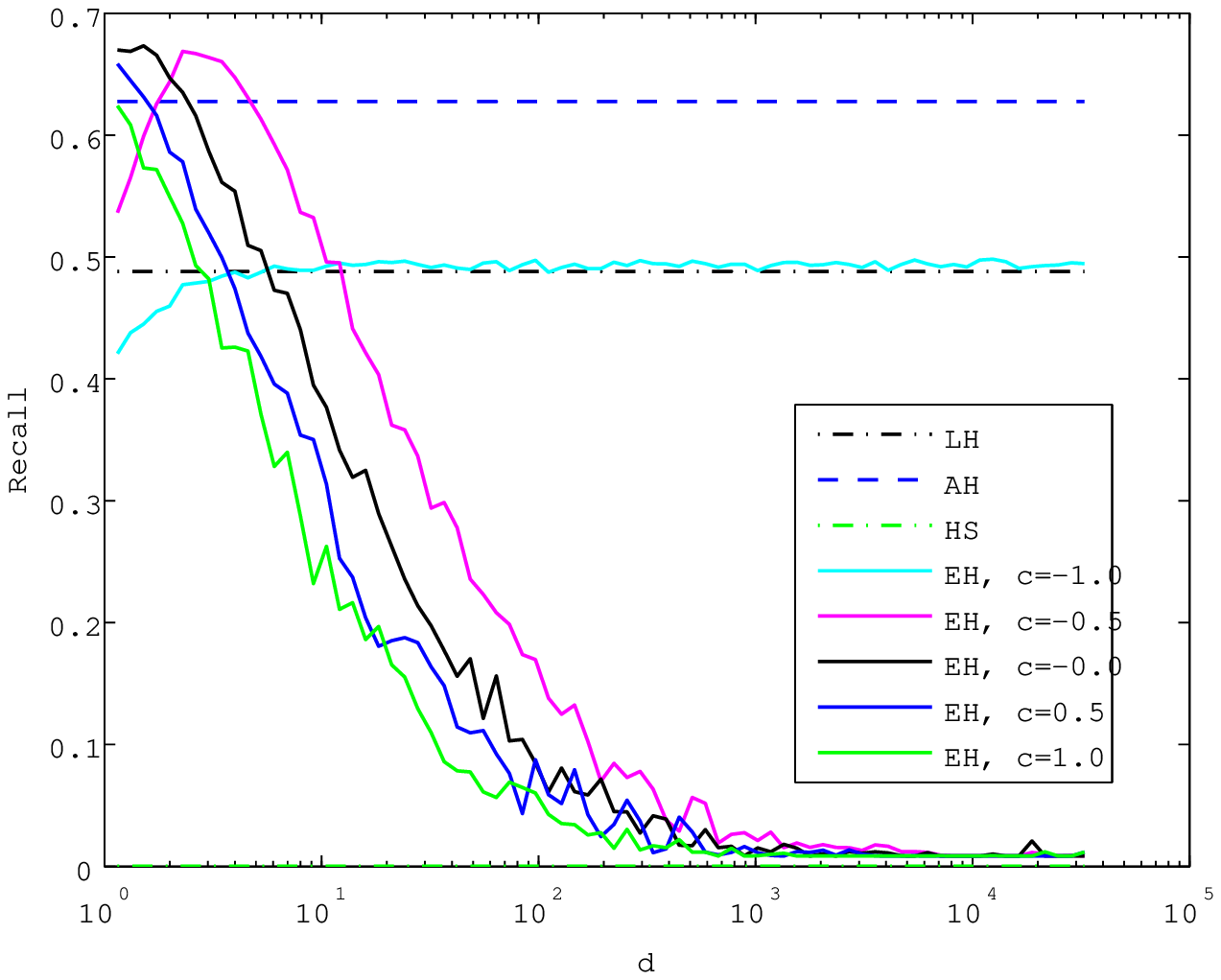}
 \end{center}
 \caption{
	Recall of each method
	on MNIST (left) and LabelMe (right).
	}
 \label{fig_RecallEcHAHSMNISTbit1024}
\end{figure*}

Figure~\ref{fig_RecallEcHAHSMNISTbit1024}
 shows the accuracy of the approximation
 of each method for MNIST and LabelMe.
From these figures,
 we can see that
 $\mathrm{Recall}(k; \mathrm{HS})$ are close to zero,
 and 
 there are regions of $d$ and $c$ where
 $\mathrm{Recall}(k; \mathrm{EH}) $ are greater than 
 $\mathrm{Recall}(k; \mathrm{LH}) $ and
 $\mathrm{Recall}(k; \mathrm{AH}) $.

\section{Discussion}
\label{sec_discussion}


The results in section~\ref{sec_accuracyExpResult}
 revealed that
 for all the data sets
 there are regions of $d$ and $c$ where
 $\mathrm{Recall}(k; \mathrm{EH}) \approx \mathrm{Recall}(k; \mathrm{LH}) $,
 $\mathrm{Recall}(k; \mathrm{EH})$ is greater than
 $\mathrm{Recall}(k; \mathrm{LH})$ and $\mathrm{Recall}(k; \mathrm{AH}) $,
 and regions where
 $\mathrm{Recall}(k; \mathrm{EH})$ is less than
 $\mathrm{Recall}(k; \mathrm{LH})$ and $\mathrm{Recall}(k; \mathrm{AH}) $.
Here, we will discuss the accuracies in the former two regions in this section.
The last one is the complement of the former two.
Moreover, we discuss the bad accuracies of $\mathrm{HS}$.


First, let us discuss the accuracies of $\mathrm{EH}$ on 
 in the regions where $\mathrm{Recall}(k; \mathrm{EH}) \approx \mathrm{Recall}(k; \mathrm{LH}) $.
When $d$ is much smaller than $r$ of the feature vectors,
 eq.(\ref{eq_inverseStereoProjection})
 is approximated as
\begin{eqnarray}
	f^{-1}(x_1, x_2, \cdots, x_N; d)
	\approx \left(
		\frac{2 x_1 d }{r^2},
		\cdots,
		\frac{2 x_N d }{r^2},
		1
	\right).
	\label{eq_inverseStereoProjectionApprox1}
\end{eqnarray}
Hence, the feature vectors are mapped to the neighborhood of the north pole of $S$.
In the case of $c=1$, since all the hyperplanes in $\tilde{V}$ cross the north pole 
 the hashing are almost similar to the hashing with hyperplanes in $V$.
Therefore,
 over the region where $d \ll r$ and $c\approx 1$
 the accuracies of $\mathrm{EH}$ is similar to the ones of $\mathrm{LH}$.


When $d$ is much greater than $r$ of the feature vectors
 the eq.(\ref{eq_inverseStereoProjection})
 is approximated as follows:
\begin{eqnarray}
	f^{-1}(x_1, x_2, \cdots, x_N; d)
	\approx \left(
		\frac{2 x_1 }{d},
		\cdots,
		\frac{2 x_N }{d},
		-1
	\right).
	\label{eq_inverseStereoProjectionApprox2}
\end{eqnarray}
Hence, the feature vectors are mapped to the neighbors
 of the south pole of $S$.
Since in the case of $c=-1$ all the hyperplanes in $\tilde{V}$ cross the south pole,
 the hashings are similar to hashing with hyperplanes in $V$.
Therefore,
 in the region where $d \gg r$ and $c\approx -1$,
 the accuracies of $\mathrm{EH}$ are similar to those of $\mathrm{LH}$.


Second, let us discuss the accuracies of $\mathrm{EH}$ 
 in the regions where $\mathrm{Recall}(k; \mathrm{EH}) > \mathrm{Recall}(k; \mathrm{LH})$.
Here, we define the optimal $c$ and $d$ as
\begin{eqnarray}
	(c_{\mathrm{opt}}, d_{\mathrm{opt}}) := \arg\max_{(c,d)}(\mathrm{Recall}(k; \mathrm{EH})).
\end{eqnarray}
From Fig.~\ref{fig_RecallEcHAHSdim512sigma1.0k100bit1024Formated}
 and Fig.~\ref{fig_RecallEcHAHSMNISTbit1024},
 we can see that
 $c_{\mathrm{opt}}$ is in $[-0.5, 0.5]$
 and does not depend greatly on the data sets.
In contrast, $d_{\mathrm{opt}}$ depends on the data sets.
In the following,
 we experimentally show that
 $d_{\mathrm{opt}}$ approximately corresponds to
 a value for which almost all feature vectors are mapped to the south hemisphere of $S$.
In other words,
 we consider that
 solution 1 in section~\ref{sec_solutionToProblems}
 improves the accuracies.

\begin{figure*}[htb]
    \includegraphics[scale=0.5]{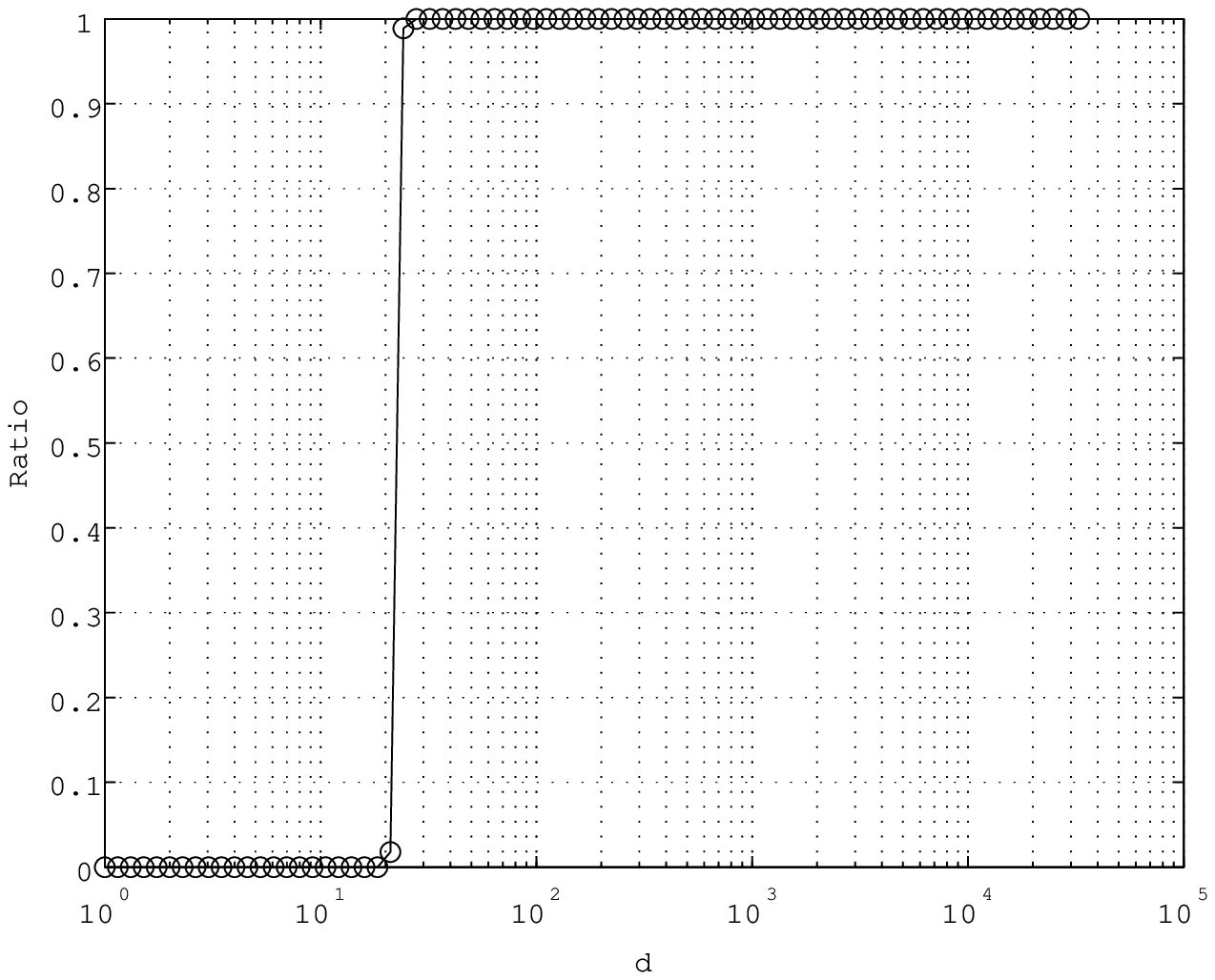}
	\includegraphics[scale=0.5]{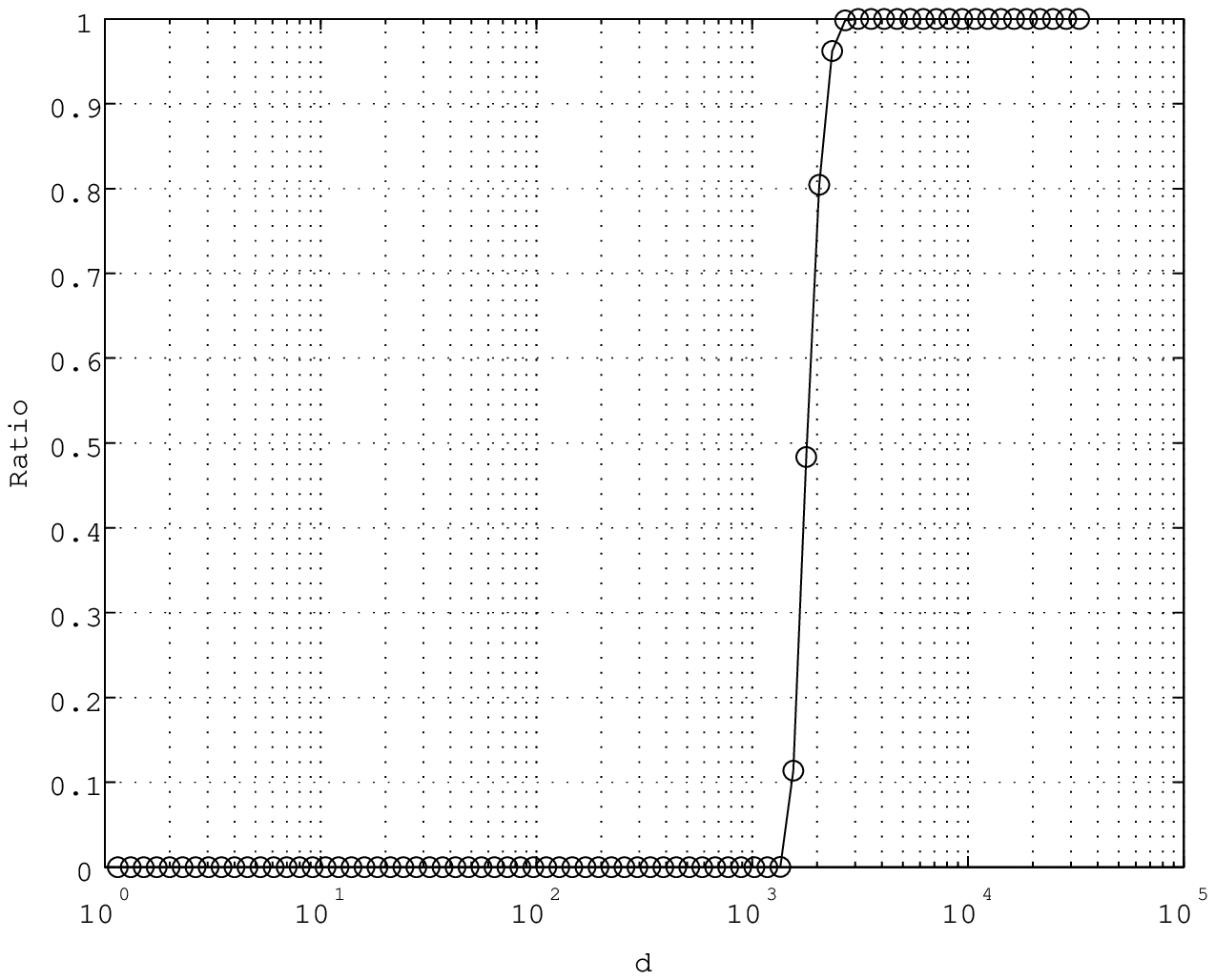}
	\includegraphics[scale=0.5]{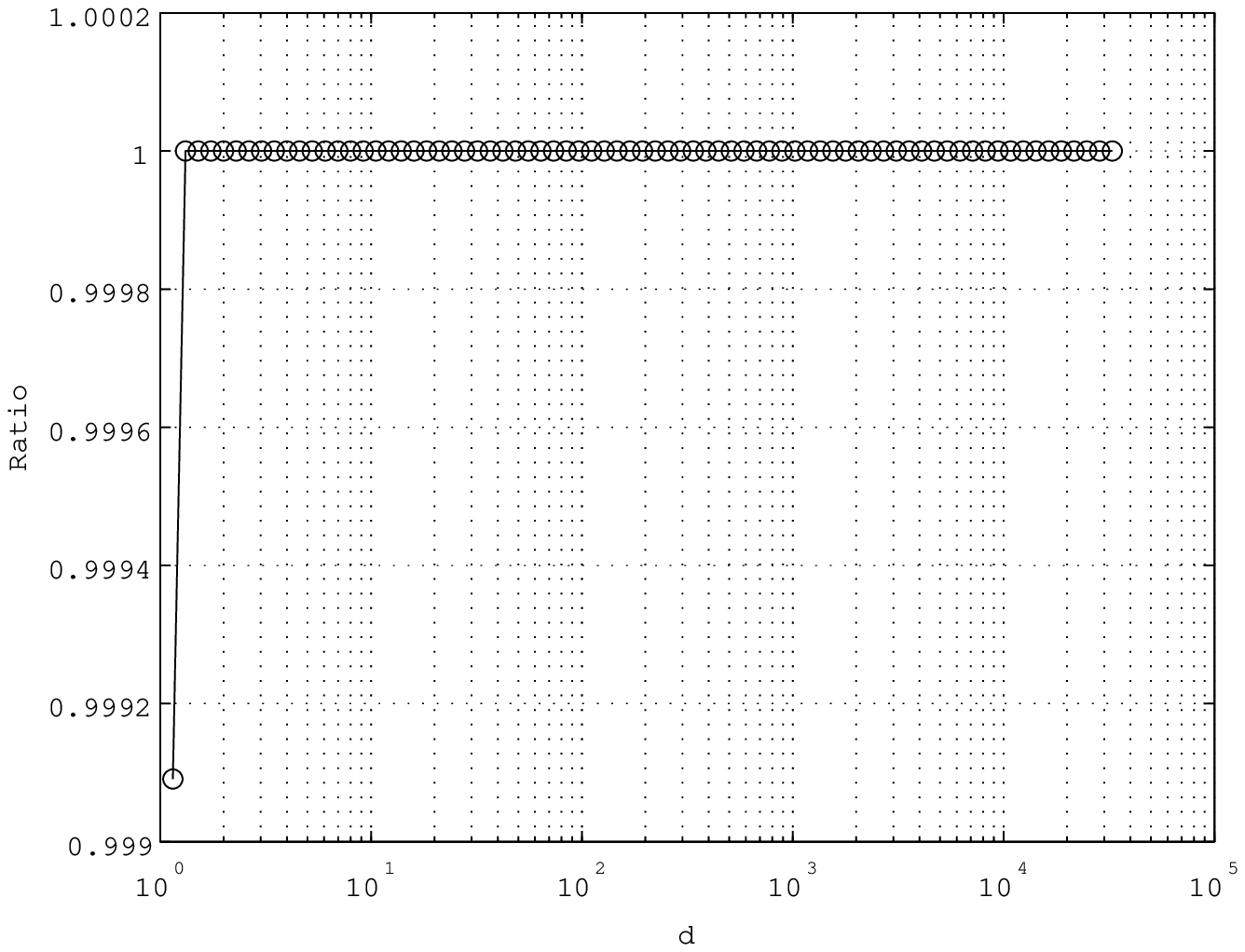}
 \caption{
	Dependency of $\mathrm{Ratio}(d)$ on $d$
	for the artificial data (upper-left),
	MNIST (upper-right)
	and LabelMe (lower-left).
	}
 \label{fig_wariai}
\end{figure*}

We calculated the ratio $\mathrm{Ratio}(d)$
 of the data mapped to the south hemisphere of $S$
 to the whole data.
Figure~\ref{fig_wariai} shows the dependency of $\mathrm{Ratio}(d)$ on $d$.


That $\mathrm{Ratio}(d)$ is almost equal to 1
 means almost all of the feature vectors are mapped to the south hemisphere.
Let $d_*$ be the value at which $\mathrm{Ratio(d)}$ becomes greater than $0.99$:
 $d_* := \arg \inf_{d}( \mathrm{Ratio}(d) > 0.99)$
Table~\ref{table_ds} shows $d_*$ and $d_{\mathrm{opt}}$ for each data set.
From the figure, we can see that
 $d_{\mathrm{opt}}$ have the same order as $d_*$ and are greater than $d_*$.
Therefore, almost all the feature vectors were mapped to the south hemisphere at $d_{\mathrm{opt}}$.
\begin{table}[htb]
	\begin{center}
	\caption{$d_*$ and $d_{\mathrm{opt}}$ }
	\begin{tabular}{l|r|r}
		\hline
		\hline
		  & $d_*$ & $d_{\mathrm{opt}}$ \\
		\hline
		Artificial data set
			& 27.86 & 32.00  \\
		MNIST
			& 2,702 & 3,104  \\
		LabelMe
			& 1.32 & 1.51  \\	
		\hline
	\end{tabular}
	\label{table_ds}
	\end{center}
\end{table}


Finally,
 we discuss the bad accuracies of $\mathrm{HS}$ on all the data sets.
For all the data sets, $\mathrm{Recall}(k; \mathrm{HS})$
 are close to zero
 and much less than $\mathrm{Recall}(k; \mathrm{EH})$ with any parameters.
We conclude that
 the deterioration in accuracy of $\mathrm{HS}$ 
 is caused not only 
 by the effects of shortcuts through the neighborhood of the infinity
 but also by wormholes.

\section{Summary and future work}
\label{sec_summary}


We proposed a new hashing scheme, called Eclipse-hashing, that leverages
 the inverse stereographic projection.
By setting the parameters properly,
 we showed that the Eclipse-hashing approximation
 is more accurate than hashing with hyperplanes.
Although hashing with hyperspheres is inaccurate in most cases
 and
 Eclipse-hashing is a kind of the hashing with hyperspheres,
 its accuracy was improved.
We think that
 the reason for this improvement is that
 it became easy to specify the cause of the performance deterioration
 by using the inverse stereographic projection.


For data sets that are different from the ones used in this paper,
 the optimal values of $\vec{C}$ and $d$ may be
 different from the ones presented here.
However, the discussion in section~\ref{sec_discussion} suggests that
 $\vec{C}$ may be near the center of $S$,
 and $d$ may be the value at which
 almost all the feature vectors are mapped to the south hemisphere.

\bibliographystyle{unsrt}
\bibliography{BibForHashing}




\end{document}